\documentclass[12pt,prd,tightenlines,nofootinbib]{revtex4}
\usepackage{bm}
\usepackage{graphics}
\usepackage{rotating}
\usepackage{epsfig}
\usepackage{slashed}
\begin{document}
\title{
Rare $\Lambda_b\to n l^+l^-$  decays in
the relativistic quark-diquark picture}  
\author{R. N. Faustov}
\author{V. O. Galkin}
\affiliation{Institute of Informatics in Education, FRC CSC RAS,
  Vavilov Street 40, 119333 Moscow, Russia}

\begin{abstract}
The form factors of the rare  $\Lambda_b\to n l^+l^-$  decays are
calculated in the framework of the relativistic quark-diquark picture of
baryons with the consistent account of the relativistic effects. Their
momentum transfer squared dependence is determined explicitly in the
whole accessible kinematical range. The decay branching fractions,
forward-backward asymmetries and the fractions of longitudinally
polarized dileptons are determined. The branching
fraction of the rare $\Lambda_b\to n\mu^+\mu^-$ decay are found to be ${\rm Br}(\Lambda_b\to n
\mu^+\mu^-)=(3.75\pm0.38)\times10^{-8}$ and thus could be measured at
the LHC. Prediction for the branching fraction of the rare radiative
$\Lambda_b\to n \gamma$ decay is also given.
\end{abstract}

\maketitle

Recently significant experimental progress has been achieved in
studying the rare decays of the $\Lambda_b$ baryon. In 2011 the CDF
Collaboration \cite{cdf} reported the first observation of the rare
$\Lambda_b\to \Lambda\mu^+\mu^-$ decay. Then the LHCb Collaboration
performed detailed angular analysis \cite{lhcb} of this decay which
allowed to extract not only the total branching fraction but also
different decay distributions and asymmetries in several momentum
transfer squared bins. This year the LHCb Collaboration observed for
the first time the suppressed decay $\Lambda_b^0\to p\pi^-\mu^+\mu^-$,
where the muons do not originate from charmonium resonances
\cite{lhcb2017}. In addition contributions from the $\Lambda_b\to\Lambda(\to p\pi^-)\mu^+\mu^-$ decays were removed by requiring $m_{p\pi^-} > 1.12$~ GeV. Such decays are mediated by the $b\to d$ transition
and thus are highly suppressed in the standard model. The measured
branching fraction of this decay is of order of $10^{-8}$. This
observation indicates that other similar rare decays, such as
$\Lambda_b\to n \mu^+\mu^-$, can be observed in the near future. 

In the recent paper \cite{lbrare} we considered the rare
$\Lambda_b\to\Lambda l^+l^-$ decays in the relativistic quark-diquark
picture of baryons \cite{barmass,sllbdecay}. The analytic expressions for the
decay form factors as the overlap integrals of the initial and final
baryon wave functions were obtained. All relativistic effects
including transformations of the baryon wave functions from rest to
the moving reference frame and contributions of intermediate negative
energy states were taken into account. This allowed us to explicitly
determine the momentum transfer squared $q^2$ dependence of the decay
form factors in the whole kinematical range without additional
assumptions and extrapolations, thus increasing reliability of the
obtained predictions. On this basis various $\Lambda_b\to\Lambda
l^+l^-$ decay observables were calculated and were found to be
consistent with detailed measurements of the LHCb Collaboration
\cite{lhcb}. In this paper we extend this analysis to the consideration
of the suppressed rare $\Lambda_b\to n l^+l^-$ and $\Lambda_b\to n \gamma$  decays.

The matrix elements of the flavour changing neutral current governing
the $b\to d$ transition between baryon states is usually parametrized
by the following set of the invariant form factors \cite{gikls}
\begin{eqnarray}
  \label{eq:ff}
\!\!\!\!\!\!  \langle n(p',s')|\bar{d} \gamma^\mu b|\Lambda_b(p,s)\rangle&=& \bar
  u_{n}(p',s')\Bigl[f_1^V(q^2)\gamma^\mu-f_2^V(q^2)i\sigma^{\mu\nu}\frac{q_\nu}{M_{\Lambda_b}}+f_3^V(q^2)\frac{q^\mu}{M_{\Lambda_b}}\Bigl]
u_{\Lambda_b}(p,s),\cr
\!\!\!\!\!\! \langle n(p',s')|\bar{d} \gamma^\mu\gamma_5 b|\Lambda_b(p,s)\rangle&=& \bar
  u_{n}(p',s')[f_1^A(q^2)\gamma^\mu-f_2^A(q^2)i\sigma^{\mu\nu}\frac{q_\nu}{M_{\Lambda_b}}+f_3^A(q^2)\frac{q^\mu}{M_{\Lambda_b}}\Bigl]
\gamma_5 u_{\Lambda_b}(p,s),\qquad \cr
\!\!\!\!\!\!\langle n(p',s') | \bar{d} i\sigma^{\mu\nu}q_\nu b | \Lambda_b(p,s) \rangle &=& \bar{u}_n(p',s') \left[  \frac{f_1^{TV}(q^2)}{M_{\Lambda_b}} \left(\gamma^\mu q^2 - q^\mu \slashed{q} \right) - f_2^{TV}(q^2) i\sigma^{\mu\nu}q_\nu  \right] u_{\Lambda_b}(p,s), \cr
\!\!\!\!\!\!\!\!\!\!\!\!\! \langle n(p',s') | \bar{d}                                                                                i\sigma^{\mu\nu}q_\nu\gamma_5 b | \Lambda_b(p,s) \rangle &=& \bar{u}_n(p',s') \left[  \frac{f_1^{TA}(q^2)}{M_{\Lambda_b}} \left(\gamma^\mu q^2 - q^\mu \slashed{q} \right) - f_2^{TA}(q^2) i\sigma^{\mu\nu}q_\nu  \right]\gamma_5 u_{\Lambda_b}(p,s), 
\end{eqnarray}
where   $u_{\Lambda_{b}}(p,s)$ and
$u_{n}(p',s')$ are the Dirac spinors of the initial $\Lambda_b$ and final $n$
baryon.

Using the relativistic quark-diquark picture of baryons with the QCD-motivated
interquark potential we obtained expressions for
these form factors as the overlap integrals of the baryon wave
functions. They are given in the Appendix of
Ref.~\cite{lbrare}. Substituting in these expressions the wave
functions obtained while considering the baryon spectroscopy
\cite{barmass} we calculate the form factors in the whole accessible
kinematical range. 

\begin{figure}
\centering
  \includegraphics[width=8cm]{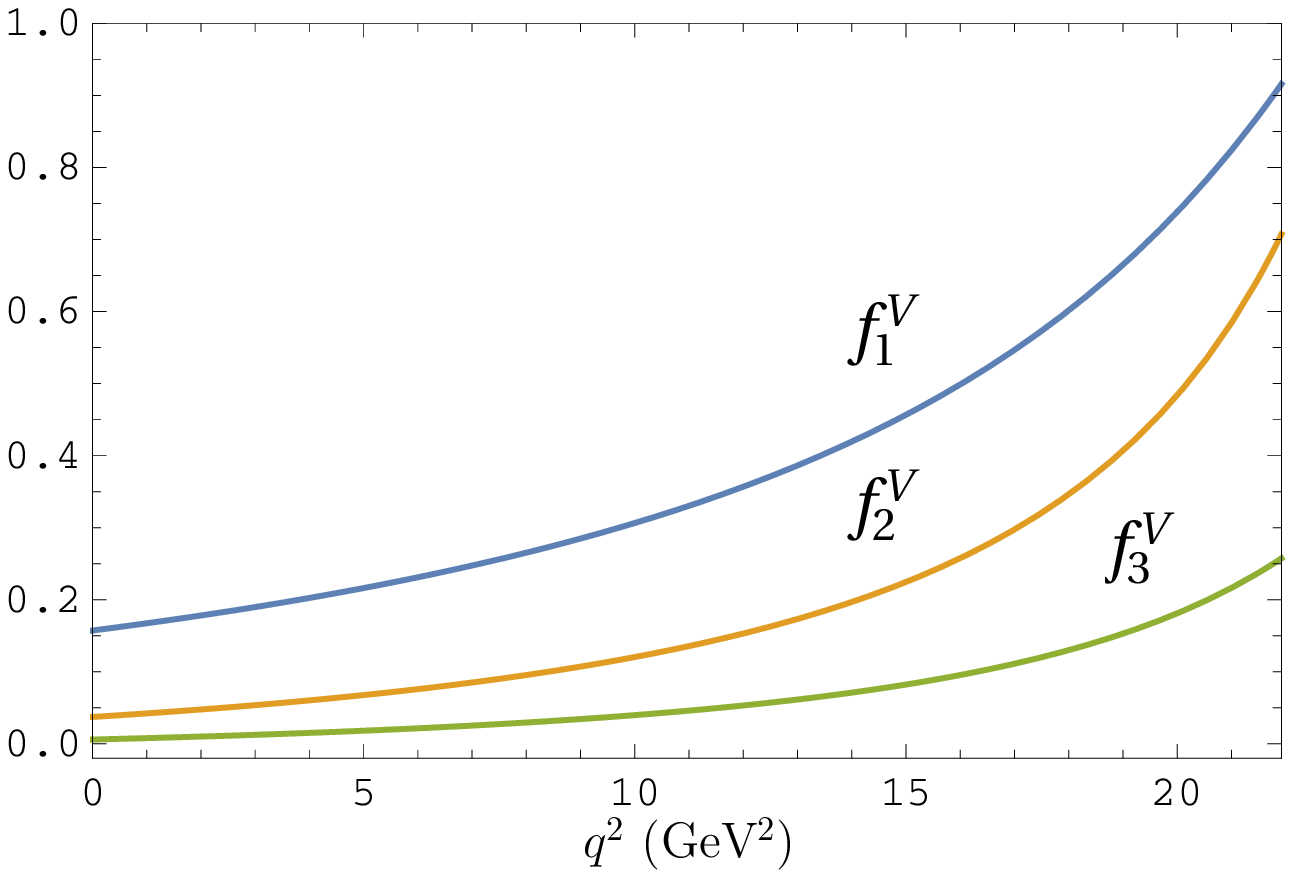}\ \
 \ \includegraphics[width=8cm]{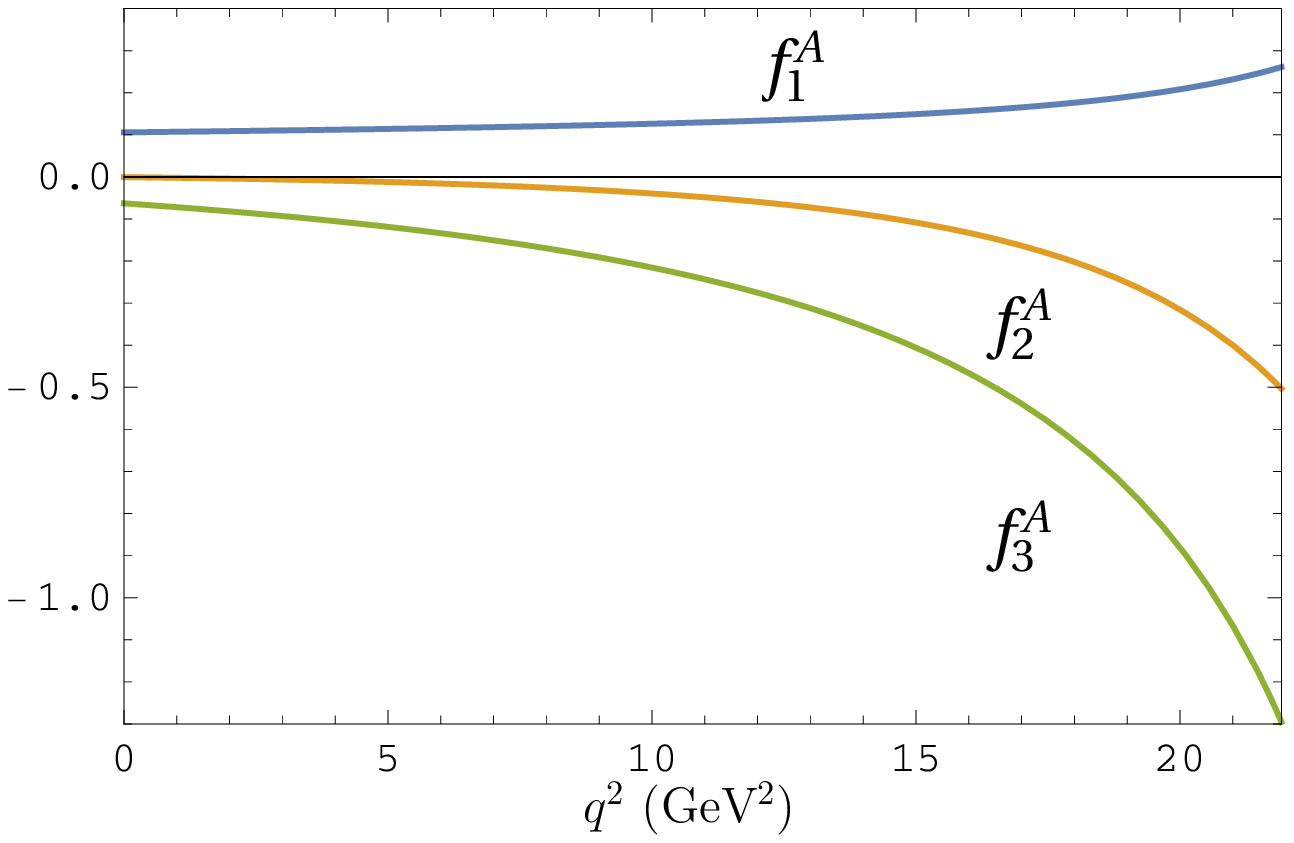}\\
\includegraphics[width=8cm]{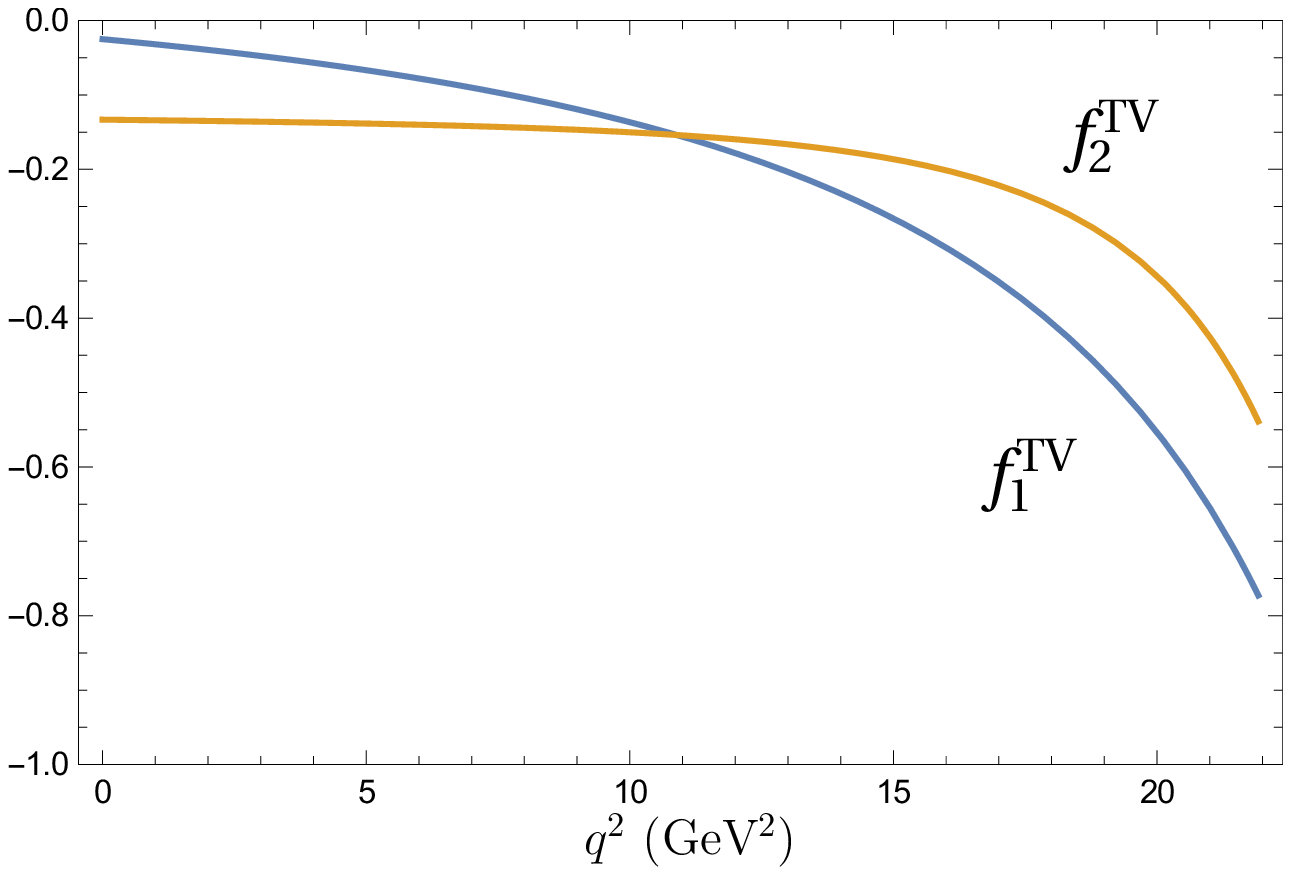}\ \
 \ \includegraphics[width=8cm]{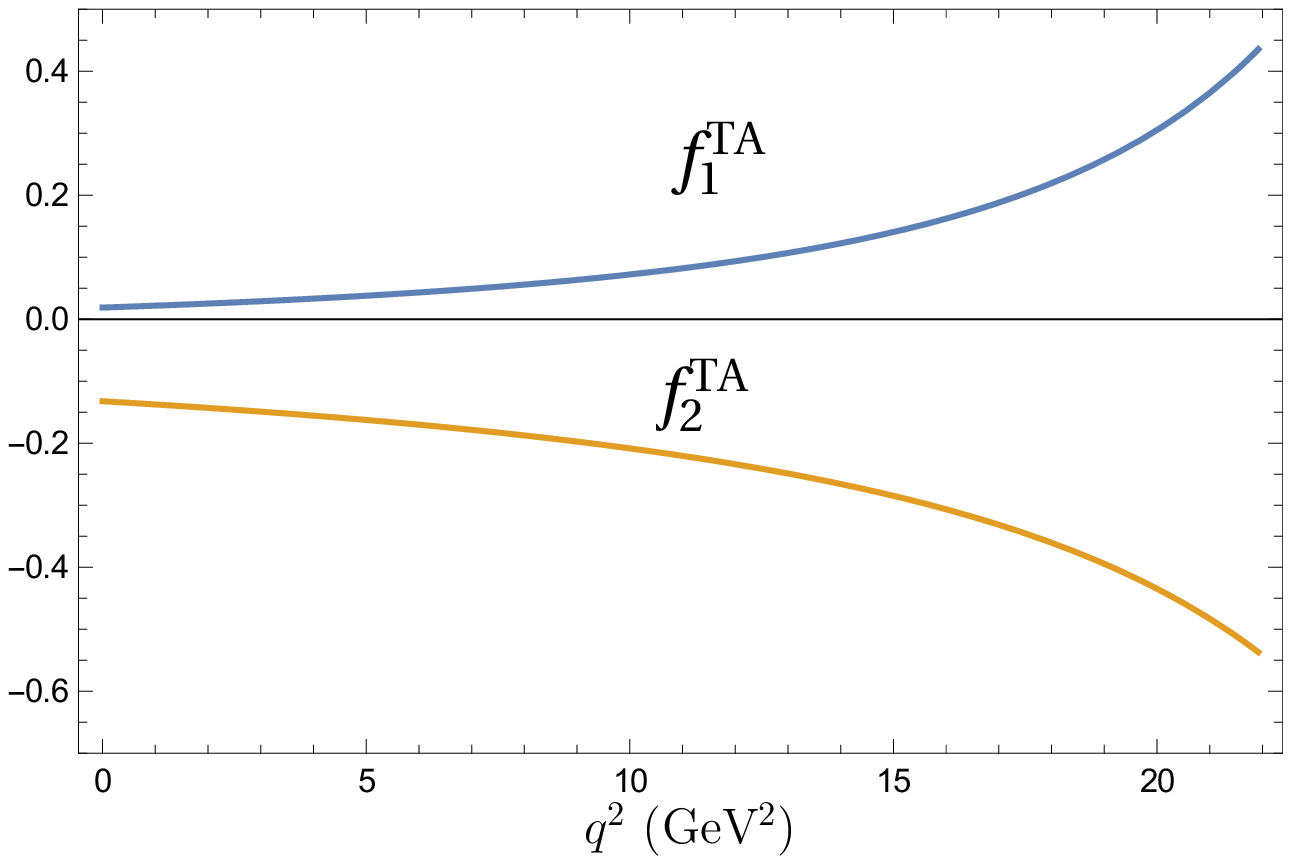}\\
\caption{Form factors of the rare $\Lambda_b\to n$ transition.    } 
\label{fig:ffLbn}
\end{figure}

We found that the numerically calculated form factors can be
approximated with high accuracy by the following analytic expression
\begin{equation}
  \label{fitff}
  F(q^2)= \frac{1}{{1-q^2/{M_{\rm pole}^2}}} \left\{ a_0 + a_1 z(q^2) +
    a_2 [z(q^2)]^2 \right\},
\end{equation}
where the variable 
\begin{equation}
z(q^2) = \frac{\sqrt{t_+-q^2}-\sqrt{t_+-t_0}}{\sqrt{t_+-q^2}+\sqrt{t_+-t_0}},
\end{equation}
$t_+=(M_B+M_\pi)^2$ and $t_0 = q^2_{\rm max} = (M_{\Lambda_b} - M_{n})^2$.  The pole
masses have the  values: $M_{\rm
  pole}\equiv M_{B^*}=5.325$ GeV for $f_{1,2}^V$, $f_{1,2}^{TV}$; $M_{\rm
  pole}\equiv M_{B_{1}}=5.723$ GeV for $f_{1,2}^A$, $f_{1,2}^{TA}$; $M_{\rm
  pole}\equiv M_{B_{0}}=5.749$ GeV for $f_{3}^V$;  $M_{\rm
  pole}\equiv M_{B}=5.280$ GeV for $f_{3}^A$.
The fitted values of the parameters $a_0$, $a_1$, $a_2$ as well as the
values of form factors at maximum $q^2=0$ and zero recoil $q^2=q^2_{\rm
  max}$ are given in Table~\ref{ffLbn}. The difference of the fitted
form factors from the calculated ones does not exceed 0.5\%. Our model form factors
are plotted in Fig.~\ref{fig:ffLbn}.

\begin{table}%[bth]
\caption{Form factors of the rare $\Lambda_b\to n$ transition. }
\label{ffLbn}
\begin{ruledtabular}
\begin{tabular}{ccccccccccc}
& $f^V_1(q^2)$ & $f^V_2(q^2)$& $f^V_3(q^2)$& $f^A_1(q^2)$ & $f^A_2(q^2)$ &$f^A_3(q^2)$& $f^{TV}_1(q^2)$ & $f^{TV}_2(q^2)$& $f^{TA}_1(q^2)$ & $f^{TA}_2(q^2)$\\
\hline
$f(0)$          &0.157 &$0.037$ & $0.007$ & 0.106 & $-0.0005$&$-0.063$&$-0.025$ & $-0.133$ & 0.019 & $-0.133$\\
$f(q^2_{\rm max})$&0.916  &$0.707$ & $0.257$ & 0.261& $-0.503$& $-1.29$ &$-0.773$ & $-0.539$ & 0.435& $-0.537$\\
$a_0$      &$0.208$&$0.160$& $0.086$& $0.086$ &$-0.166$&  $-0.277$&$-0.175$& $-0.122$& $0.144$ &$-0.178$\\
$a_1$      &$0.237$&$-0.298$&$-0.257$&$-0.124$&$0.812$&$0.448$&$0.161$&$0.449$&$-0.450$&$0.147$\\
$a_2$      &$-1.176$&$-0.224$& $0.044$& $0.550$ &$-0.939$&  $0.595$&$0.883$& $-1.45$& $0.218$ &$0.030$\\
\end{tabular}
\end{ruledtabular}
\end{table}

Now we can use the obtained form factors for the calculation of the
rare $\Lambda_b\to n l^+l^-$  decay observables.

The effective Hamiltonian for the $b\to d l^+ l^-$
transitions, taking into account the unitarity of the
Cabibbo-Kobayashi-Maskawa (CKM) matrix  takes \cite{bhi} the following form 
\begin{equation}
  \label{eq:heff}
  {\cal H}_{\rm eff} =-\frac{4G_F}{\sqrt{2}}\left[V_{td}^*V_{tb}\sum_{i=1}^{10}c_i{\cal O}_i+ V_{ud}^*V_{ub}\sum_{i=1}^{2}c_i\left({\cal O}_i-{\cal O}_i^{(u)}\right) \right],
\end{equation}
where $G_F$ is the Fermi constant, $V_{tj}$ and $V_{uj}$ are the
CKM matrix elements, $c_i$ are the Wilson coefficients
and ${\cal O}_i({\cal O}_i^{(u)})$ represent the four-quark operator
basis. Then the resulting matrix element of the transition amplitude
between baryon states can be written as \cite{pirare}
  \begin{eqnarray}
  \label{eq:mtl}
  {\cal M}&=&\frac{G_F\alpha}{\sqrt{2}\pi}
  |V_{td}^*V_{tb}|\Biggl\{\langle  n|\left[c_9^{\rm eff}\gamma_\mu(1-\gamma_5)-
    \frac{2m_b}{q^2}c_7^{\rm eff}i\sigma_{\mu\nu}q^\nu(1+\gamma_5)\right]|\Lambda_b\rangle(\bar
    l\gamma^\mu l)\cr
&&+c_{10}\langle n|\gamma_\mu(1-\gamma_5)|\Lambda_b\rangle
    (\bar l\gamma^\mu \gamma_5l)\Biggr\},
\end{eqnarray}
where the values of the Wilson coefficients $c_i$ and of the effective
Wilson coefficient $c_7^{\rm eff}$  are taken from Ref.~\cite{wc}. The 
effective  Wilson coefficient $ c_9^{\rm eff}$ contains additional
perturbative and long-distance contributions coming from hadron resonances
\begin{eqnarray}
  \label{eq:c9}
 c_9^{\rm eff}&=&c_9+h^{\rm eff}\left(\frac{m_c}{m_b},\frac{q^2}{m_b^2}\right)c_0+\lambda_u\left[h^{\rm eff}\left(\frac{m_c}{m_b},\frac{q^2}{m_b^2}\right)-h^{\rm eff}\left(\frac{m_u}{m_b},\frac{q^2}{m_b^2}\right)\right](3c_1+c_2)\cr
&&-
\frac12 h\left(1,\frac{q^2}{m_b^2}\right)(4c_3+4c_4+3c_5+c_6)-\frac12
h\left(0,\frac{q^2}{m_b^2}\right)(c_3+3c_4)\cr
&&+\frac29(3c_3+c_4+3c_5+c_6).
\end{eqnarray}   
Here $\lambda_u=\frac{V_{ud}^*V_{ub}}{V_{td}^*V_{tb}}$,  the
coefficient $c_0=3c_1+c_2+3c_3+c_4+3c_5+c_6$ and
\begin{eqnarray*} 
h\left(\frac{m_c}{m_b},  \frac{q^2}{m_b}\right) & = & 
- \frac{8}{9}\ln\frac{m_c}{m_b} +
\frac{8}{27} + \frac{4}{9} x 
-  \frac{2}{9} (2+x) |1-x|^{1/2} \left\{
\begin{array}{ll}
 \ln\left| \frac{\sqrt{1-x} + 1}{\sqrt{1-x} - 1}\right| - i\pi, &
 x \equiv \frac{4 m_c^2}{ q^2} < 1,  \\
 & \\
2 \arctan \frac{1}{\sqrt{x-1}}, & x \equiv \frac
{4 m_c^2}{ q^2} > 1,
\end{array}
\right. \\
h\left(0, \frac{q^2}{m_b} \right) & = & \frac{8}{27} - 
\frac{4}{9} \ln\frac{q^2}{m_b} + \frac{4}{9} i\pi.
\end{eqnarray*}
The function
\begin{equation}
  \label{eq:c9np}
h^{\rm
  eff}\left(\frac{m_c}{m_b},\frac{q^2}{m_b^2}\right)=h\left(\frac{m_c}{m_b},\frac{q^2}{m_b^2}\right)+
\frac{3\pi}{\alpha^2 c_0} \sum_{V_i=J/\psi,\psi(2S)\dots}\frac{\Gamma(V_i\to l^+l^-)M_{V_i} }{M_{V_i}^2-q^2-iM_{V_i}\Gamma_{V_i}}
\end{equation}
contains additional long-distance contributions originating from the $c\bar c$ resonances [$J/\psi,
\psi(2S)\dots$]. In our study we include contributions of the vector $V_i(1^{--})$ charmonium states: $J/\psi$,
$\psi(2S)$, $\psi(3770)$, $\psi(4040)$, $\psi(4160)$ and $\psi(4415)$,
with their masses ($M_{V_i}$), leptonic [$\Gamma(V_i\to l^+l^-)$] and
total ($\Gamma_{V_i}$) decay widths taken from PDG \cite{pdg}. Similar expression holds
for the function $h^{\rm
  eff}\left(\frac{m_u}{m_b},\frac{q^2}{m_b^2}\right)$, where the
long-distance contributions now come from $\rho$ and
$\omega$ states. 

\begin{figure}
  \centering
 \includegraphics[width=8cm]{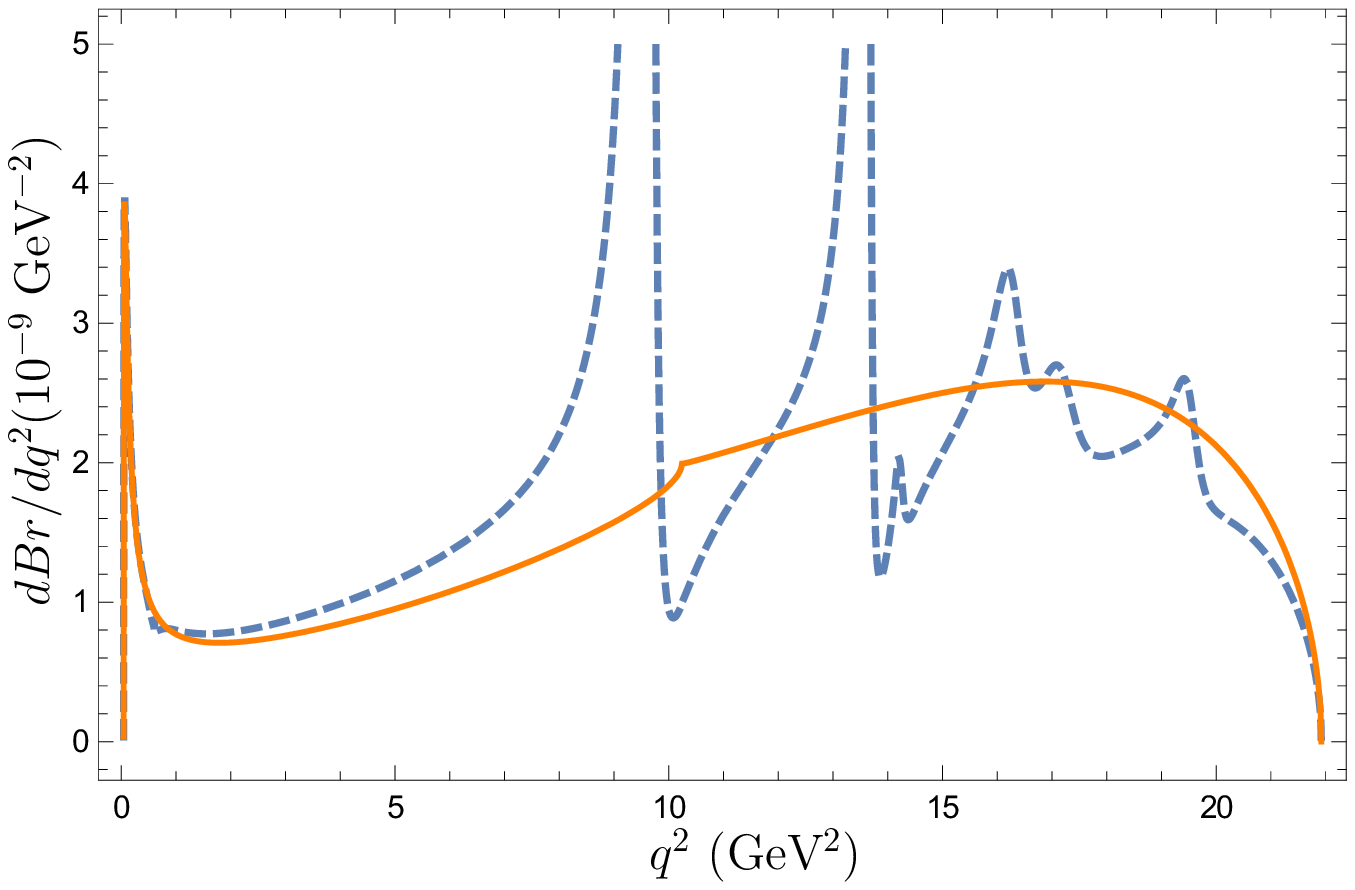}\ \
 \  \includegraphics[width=8cm]{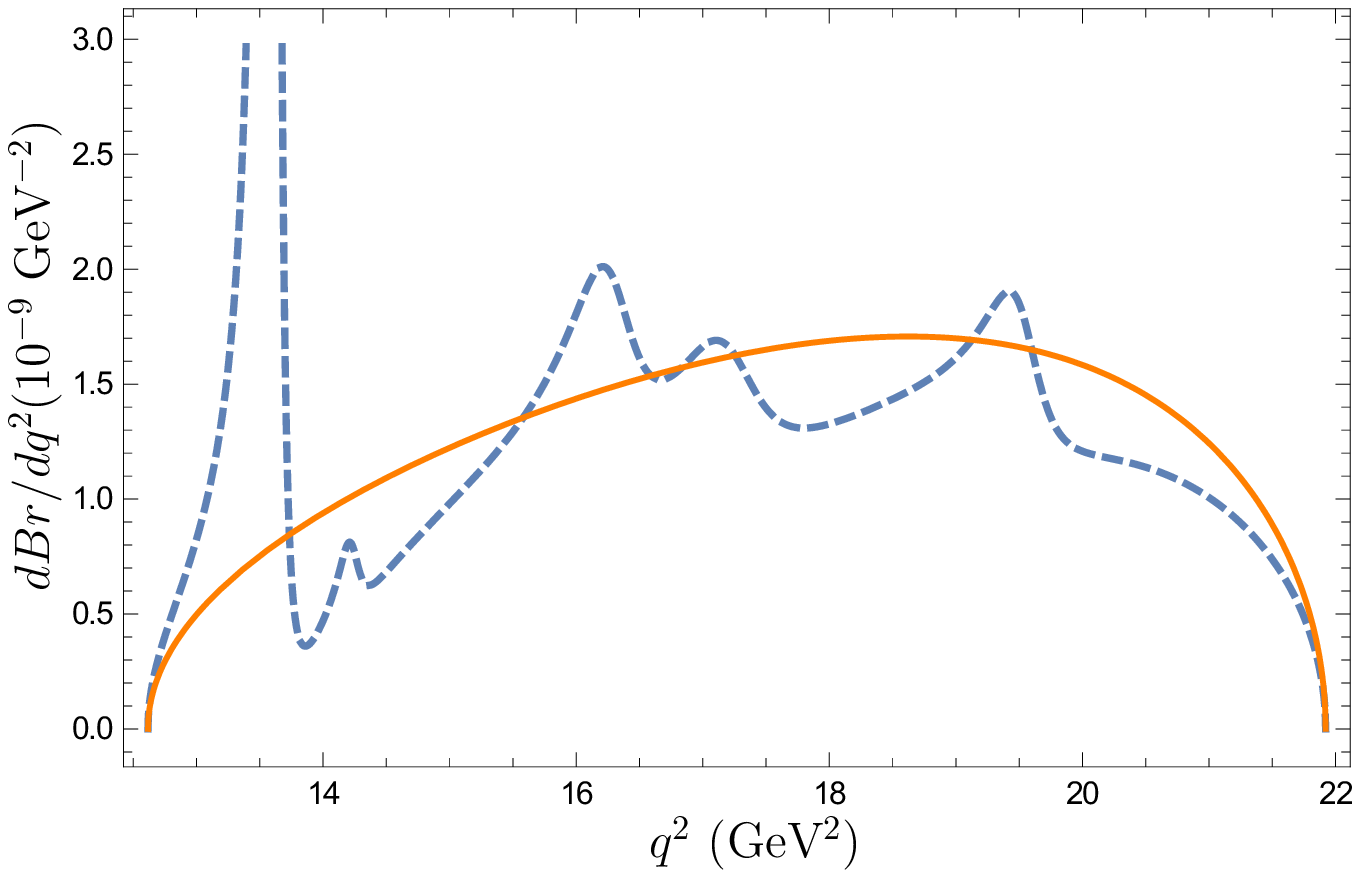}
  \caption{The differential branching fractions for the
    $\Lambda_b\to n\mu^+\mu^-$ (left) and $\Lambda_b\to n\tau^+\tau^-$ (right)
    rare decays. }
  \label{fig:brLbn}
\end{figure}

\begin{figure}
  \centering
 \includegraphics[width=8cm]{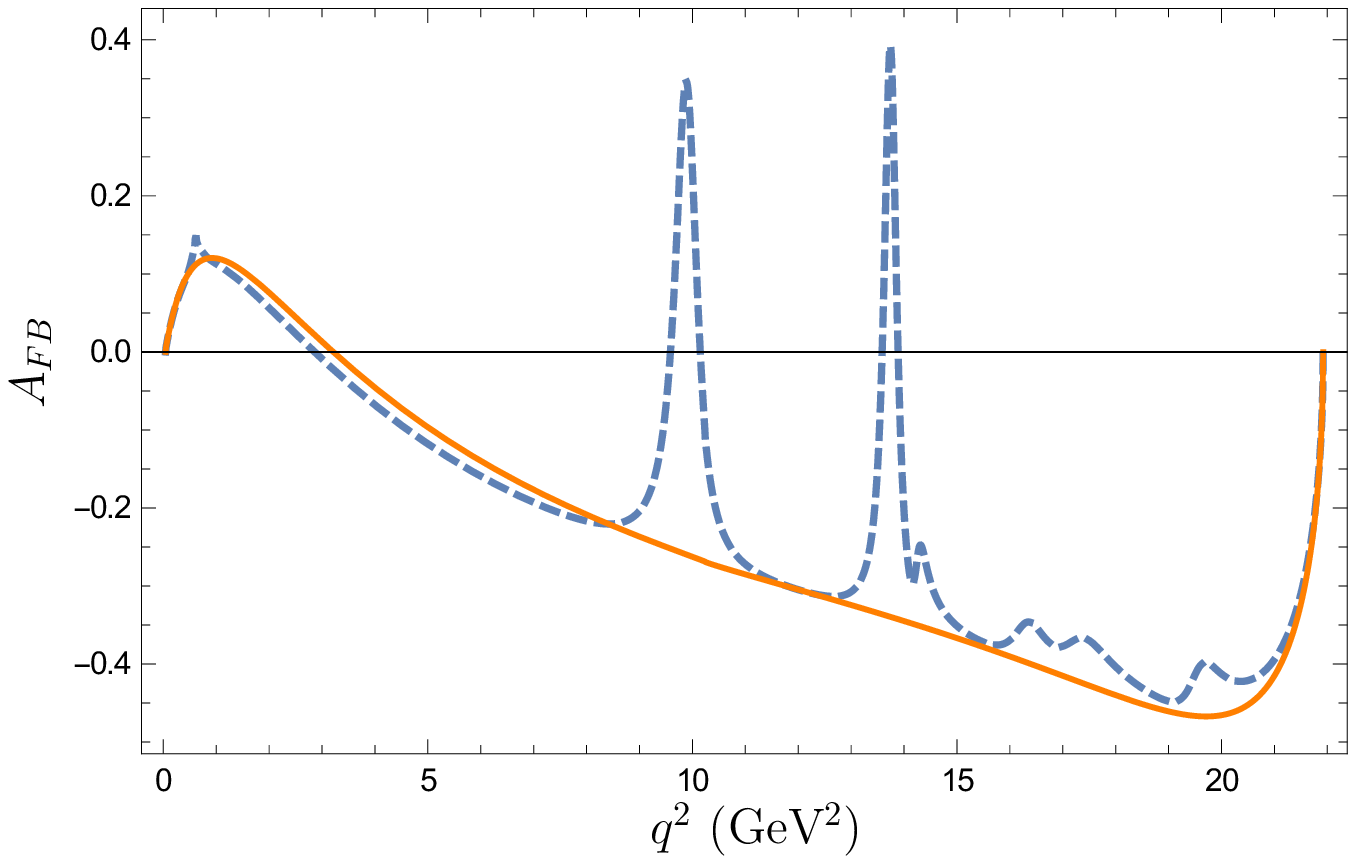}\ \
 \  \includegraphics[width=8cm]{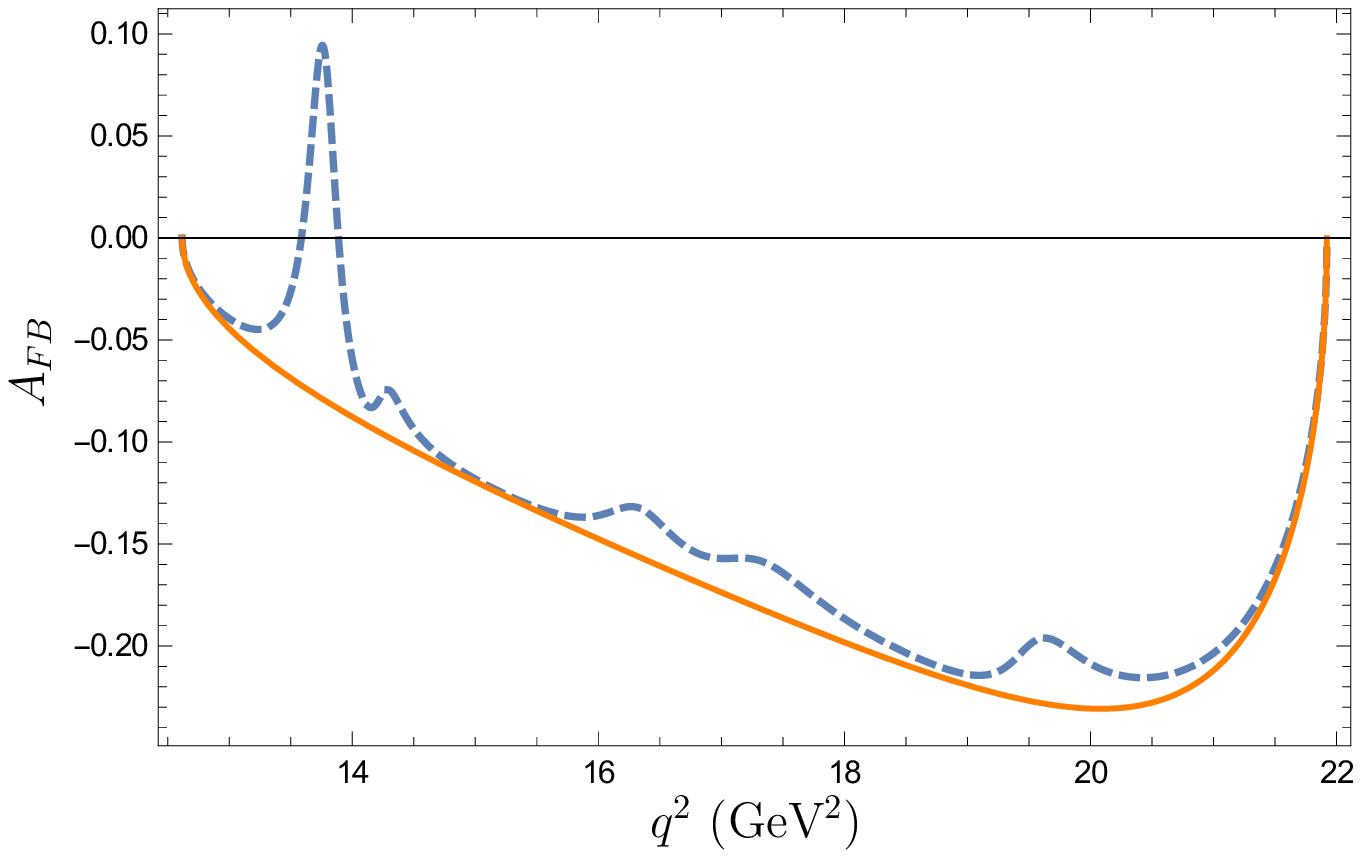}
\caption{The 
    forward-backward asymmetries  $A_{FB}(q^2)$ in the
    $\Lambda_b\to n \mu^+\mu^-$ (left) and $\Lambda_b\to n
    \tau^+\tau^-$ (right) rare decays.}
  \label{fig:afb}
\end{figure}

\begin{figure}
  \centering
 \includegraphics[width=8cm]{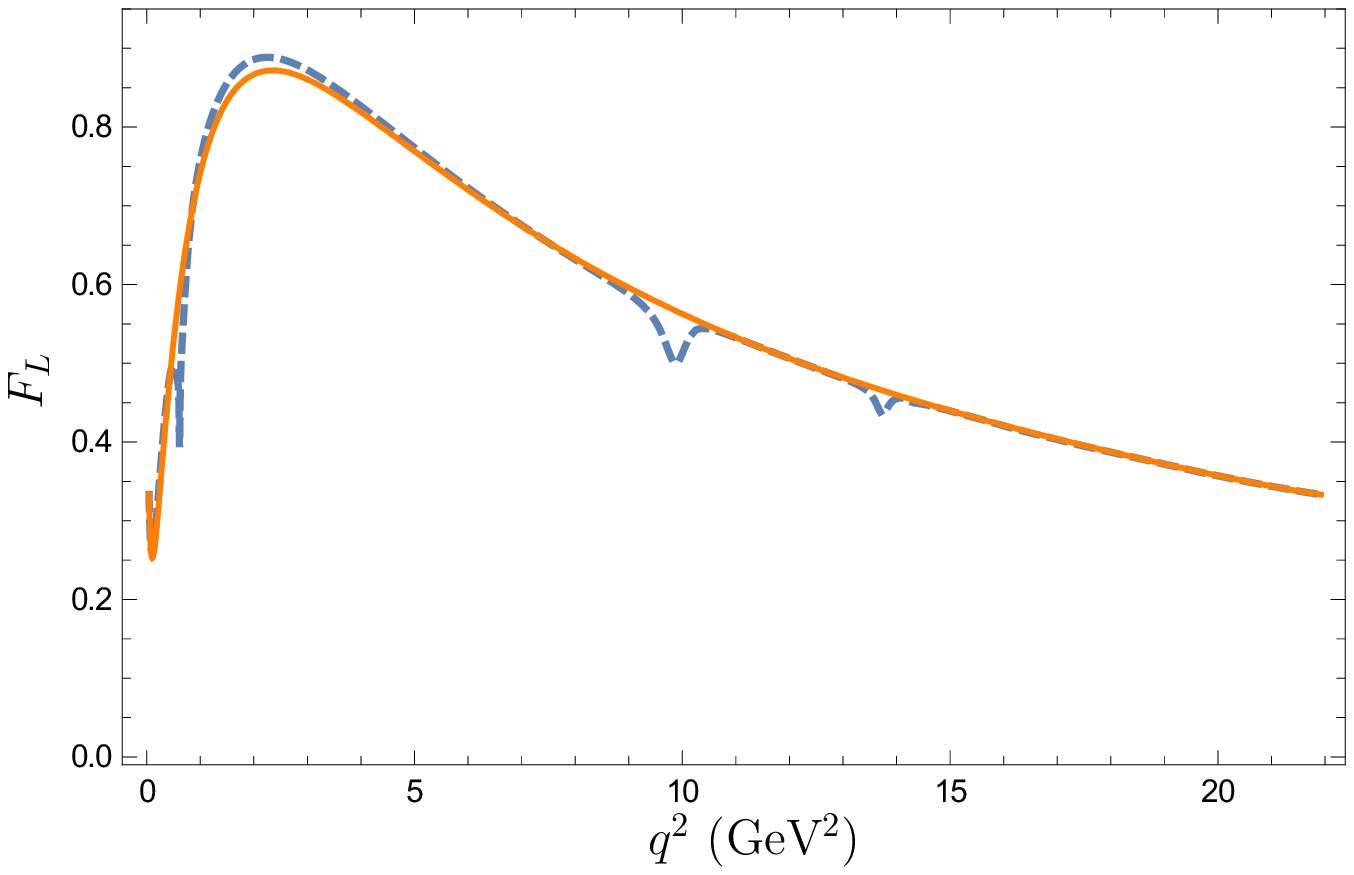}\ \
 \  \includegraphics[width=8cm]{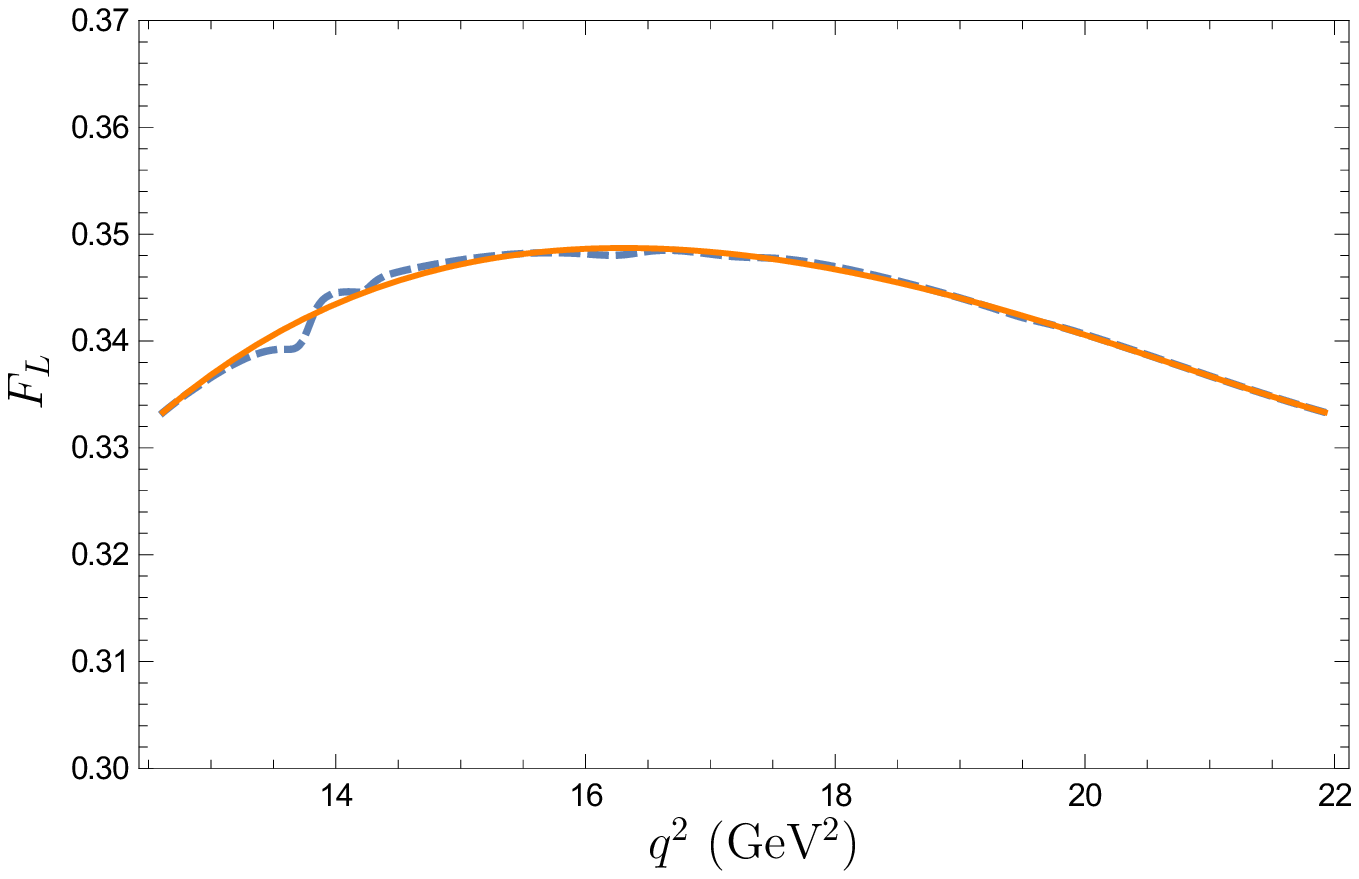}

  \caption{The fraction of longitudinally polarized
    dileptons $F_L(q^2)$  in the  $\Lambda_b\to n \mu^+\mu^-$ (left) and $\Lambda_b\to n
    \tau^+\tau^-$ (right)
    rare decays. }
  \label{fig:fl}
\end{figure}

The differential decay distribution is given by
\begin{equation}
  \label{eq:ddGl}
\!\!\!\!\!\!  \frac{d^2 \Gamma(\Lambda_b\to n l^+l^-)}{d q^2d\cos\theta}=\frac{d
    \Gamma(\Lambda_b\to n l^+l^-)}{d q^2}\left\{\frac38(1+\cos^2\theta)[1-F_L(q^2)]+ A_{FB}(q^2)\cos\theta+\frac34F_L(q^2)\sin^2\theta\right\},\qquad
\end{equation}
where $\theta$ is the angle between the $\Lambda_b$ baryon and the positively
charged lepton in the dilepton rest frame, 
$$A_{FB}(q^2)=\frac{\int^1_0\frac{d^2\Gamma}{dq^2d\cos\theta}d\cos\theta-\int^0_{-1}\frac{d^2\Gamma}{dq^2d\cos\theta}d\cos\theta}{d\Gamma/dq^2},$$ 
is the forward-backward asymmetry and $F_L(q^2)$ is
the fraction of longitudinally polarized dileptons. The explicit
expressions for the differential branching fractions and asymmetries
in terms of the form factors are given in
Ref.~\cite{lbrare}. Note that the rare decays $\Lambda_b\to
n l^+l^-$ are additionally suppressed by the ratio of the CKM
matrix elements $(|V_{td}|/|V_{ts}|)^2$ with respect to the
$\Lambda_b\to \Lambda l^+l^-$ decays.  Substituting in these
expressions the calculated form
factors we get predictions for the differential decay branching fractions,
forward-backward asymmetries  $A_{FB}(q^2)$ and the fractions of
longitudinally polarized dileptons $F_L(q^2)$. They are plotted in
Figs.~\ref{fig:brLbn}--\ref{fig:fl} for the $\Lambda_b\to n\mu^+\mu^-$
and $\Lambda_b\to n\tau^+\tau^-$ rare decays. By solid and dashed lines we plot
theoretical results obtained without and with inclusion of the
long-distance contributions to the Wilson coefficient $c_9^{\rm eff}$. The values of the total branching
fractions and averaged forward-backward asymmetries  $\langle A_{FB}(q^2)\rangle$ and fractions of
longitudinally polarized dileptons $\langle F_L(q^2)\rangle$ are given
in Table~\ref{brass}. The branching fractions were calculated without
inclusion of the hadron resonance contributions, while  $\langle
A_{FB}(q^2)\rangle$ and $\langle F_L(q^2)\rangle$ are presented both
without (nonres.) and with (res.) their inclusion. We estimate the
theoretical errors of our predictions, which emerge from the
uncertainties in the calculation of the decay form factors, to be
about 10\%.   

\begin{table}%[bth]
\caption{Predictions for the branching fractions, averaged rare
  decay forward-backward asymmetries and polarization fractions. }
\label{brass}
\begin{ruledtabular}
\begin{tabular}{cccccc}

Decay&Br& \multicolumn{2}{c}{$\langle A_{FB}\rangle$}& \multicolumn{2}{c}{$\langle F_L\rangle$}\\
&$(\times10^{-8})$& nonres.& res.  & nonres.& res. \\
\hline
$\Lambda_b\to n e^+e^-$&3.81&$-0.294$&$-0.301$&0.499&0.543\\
$\Lambda_b\to n \mu^+\mu^-$&3.75&$-0.298$&$-0.299$&0.504&0.555\\
$\Lambda_b\to n \tau^+\tau^-$&1.21&$-0.173$&$-0.148$&0.344&0.339\\
\end{tabular}
\end{ruledtabular}
\end{table}

Using calculated values of the rare decay form factors we can also
predict the
exclusive rare radiative $\Lambda_b\to n\gamma$ decay rate. It  is
expressed in terms of the decay form factors by
\begin{equation}
\label{rad}
\Gamma(\Lambda_b\to n\gamma)=
\frac{\alpha }{64\pi^4} G_F^2m_b^2M_{\Lambda_b}^3|V_{tb}V_{td}|^2
|c_7^{\rm eff}(m_b)|^2 (|f_{2}^{TV}(0)|^2+|f_{2}^{TA}(0)|^2) 
\left(1-\frac{M_{n}^2}{M_{\Lambda_b}^2} \right)^3.
\end{equation}
Substituting their calculated values we get the prediction for the
branching fraction
\begin{equation}
  \label{eq:rbr}
  {\rm Br}(\Lambda_b\to n\gamma)=3.7\times 10^{-7}.
\end{equation}

In this paper we calculated the form factors of the rare $b\to d$
transitions between baryon states in the relativistic quark-diquark
picture of baryons. These form factors were explicitly determined in
the whole accessible kinimatical range without extrapolations with the
account of the relativistic effects including wave function
transformations from rest to the moving reference frame and
intermediate contributions of the negative energy states. On this
basis predictions for the branching fractions of the rare semileptonic $\Lambda_b\to n l^+l^-$
and rare radiative $\Lambda_b\to n\gamma$ decays were obtained. The rare semileptonic
branching fractions were found to be of order of $10^{-8}$, while the rare
radiative decay is predicted to have branching fraction of order of
$10^{-7}$. Thus the CKM suppressed rare baryon decays governed by the $b\to d$ transition could be accessible for observation at LHC by the
LHCb Collaboration. The predictions for the averaged values of the
forward-backward asymmetry $\langle A_{FB}(q^2)\rangle$ and the fractions of
longitudinally polarized dileptons $\langle F_L(q^2)\rangle$ are also given.

\end{document}